\documentclass[12pt,onecolumn,twoside]{article}
\usepackage{color,graphics,verbatim,graphicx}
\usepackage{sidecap}
\begin{document}

\title{\Huge\bf Tidal effects and periastron events in binary stars}
\vskip5cm
\author{Gloria Koenigsberger \& Edmundo Moreno\\ 
Universidad Nacional Aut\'onoma de M\'exico\\
{\it gloria@fis.unam.mx; edmundo@astroscu.unam.mx}}
\maketitle


\vskip1cm
\centerline{\Large\bf ABSTRACT}
\vskip0.3cm
\noindent{\large Binary stars in eccentric orbits are frequently reported to present increasing levels
of activity around periastron passage.  In this paper\footnote{Based on the poster presented at {\it Hot Massive
Stars...A lifetime of Influence}, Workshop held at Lowell Observatory, 2008  October 12--15, celebrating 
Peter Conti's birthday} we present results of a calculation from first principles
of the velocity field on the surface of a star that is perturbed by a binary companion.  This allows 
us to follow the orbital phase-dependence of the amount of kinetic energy that may be dissipated through 
the viscous shear, $\dot{E}$,  driven by tidal interactions.  For stars with relatively small stellar radii compared
with the orbital separation at periastron ($R_1/r_{per}\leq$ 0.14), a clear maximum occurs before periastron
passage for sub-synchronous rotation and after periastron for super-synchronous rotation. For larger stellar 
radii however, $\dot{E}$ oscillates  over the orbital cycle and periastron passage does not cause a particularly
greater enhancement in energy dissipation rates than some of the other orbital phases. 
Finally, we perform exploratory calculations for a WR/LBV binary system that in 1993-1994 underwent an
LBV-like eruption, HD 5980.  Our $\dot{E}$ calculations reproduce  the oscillations that appear around periastron 
passage in HD5980's recent visual light curve.  We suggest that the energy dissipation from tidal flows 
in asynchronously rotating binary stars may provide  a mechanism by which evolving stars may be driven into 
an active state.  Given the nature of the tidal perturbations, the resulting mass-loss distribution is
expected to be non-uniform over the stellar surface and highly time-dependent.
 }

\vskip1cm

{\large
\section{Introduction}
A number of binary systems show evidence of enhanced activity around periastron passage.
Among these,  $\eta$ Car is the most extreme
and best documented example of periodic brightening at X-ray, visual and IR wavebands 
associated with periastron passages.  Recently, van Genderen \& Sterken\footnote{2007, IBVS, 5782} 
suggested that these periastron events  may have the same physical cause as the milder 
``periastron effects"  exhibited by many less renowned  eccentric binaries in which small 
enhancement ($\Delta$m$_v\sim$0.01--0.03$^{mag}$) in the visual brightness of the system 
around periastron passage are observed.  They suggested that the fundamental cause of the 
effects may reside in the enhanced tidal force present during periastron passage. 

Our study of tidal interactions  and their potential role in sparking stellar
activity was first inspired by the peculiar behavior of the WR/LBV system 
HD 5980.\footnote{Koenigsberger, G. 2004, RMAA,40, 107; Georgiev,
Hillier \& Koenigsberger, this meeting.}  In particular, the orbital phase-dependent
behavior of the line-profile variations led us to raise the question of whether the 
periastron passage could affect the intrinsic stellar wind properties, 
possibly enhancing the mass-loss rate around this phase.\footnote{Koenigsberger et al. 2002,
ASPCS 260, 507}.  Our objective is to explore whether the mechanism involved in causing the 
instabilities is the  dissipation  of kinetic energy  of the viscous  flows  that are driven 
by the tidal interactions. 

\begin{SCfigure}   
\includegraphics[width=0.65\columnwidth]{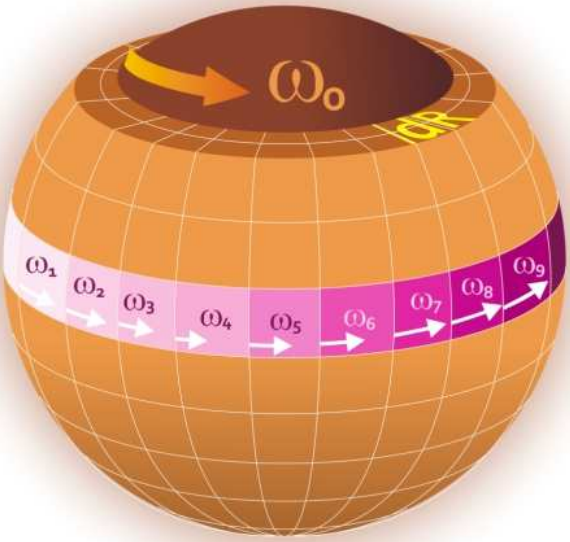}
\caption{Artistic representation of the one-layer model for the tidal interaction
calculations.  The inner body of the star is assumed to rotate uniformly,
while the perturbations acting on the surface layer produce a velocity
field leading to local angular velocities that differ from the underlying uniform rotation.
}
\end{SCfigure}


\section{Method for $\dot{E}$ calculation}

A system is in synchronous rotation when the orbital angular velocity $\Omega$ equals the angular
velocity of rotation $\omega_0$.  In eccentric orbits, the degree of synchronicity varies with orbital
phase.  We use periastron passage as the reference point for defining the synchronicity
parameter $\beta_{per}=\omega_0/\Omega_{per}= 0.02 P \frac{v_{rot}(1-e)^{3/2}}{R_1(1+e)^{1/2}}$, 
where $e$ is the orbital eccentricity, the rotation velocity $v_{rot}$ is given in km/s, 
the orbital period P is given in days, and the stellar equilibrium radius R$_1$ is given in solar units. 
When at any orbital phase $\beta\neq$1, non-radial oscillations are excited, driven by the tidal interactions.  
We refer to the azimuthal component of the forced oscillations as ``tidal flows".

Our method consists of computing  the motion of a Lagrangian grid of surface elements distributed along
a series of parallels (i.e., rings with different polar angles) covering the  surface of the star with 
mass M$_1$ as it is perturbed by its companion of mass M$_2$.  
The main stellar body below the perturbed layer is assumed to have uniform rotation. The equations of 
motion that are solved for the set of surface elements include the gravitational fields of M$_1$ and M$_2$, the 
Coriolis force, and gas pressure. The motions of all surface elements  are coupled through the  viscous 
stresses included in the equations of motion. The kinetic energy of the tides may be dissipated through 
viscous shear, thus leading to an energy dissipation rate, $\dot{E}$.  The rate of energy dissipation per 
unit volume is given by the matrix product $\dot{E} = - {\bf P_{\eta}}:\nabla \mbox{\boldmath$v$}$, 
where ${\bf P_{\eta}}$ is the viscous part of the stress tensor and {\boldmath $v$} is the velocity of 
a volume element with respect to the center of the star.\footnote{For details see Moreno \& Koenigsberger 
1999, RMAA, 35, 157; Moreno et al. 2005, A\&A, 437, 641; and  Toledano et al. 2007, A\&A, 458, 413}.

The benefits of our model are: 1) we  make no {\it a priori} assumption regarding the mathematical 
formulation of the tidal flow structure since we derive the velocity field {\boldmath $v$} from  
first principles; 2) the method is not limited to slow stellar rotation rates nor to small orbital 
eccentricities; and 3) it is computationally inexpensive.

\begin{SCfigure}   
\includegraphics[width=0.65\columnwidth]{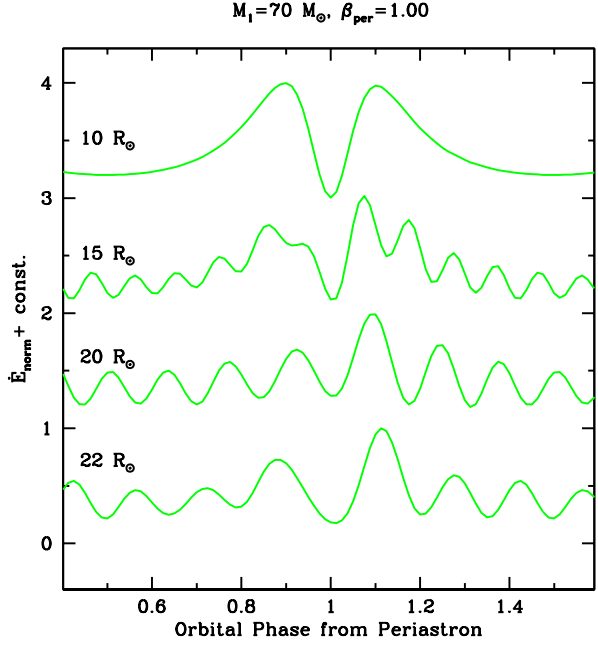}
\caption{Energy dissipation rates plotted as a function of orbital phase  
for a  M$_1=$70 M$_\odot$ primary star of different stellar radii.  In all cases, the
star is in synchronous rotation at periastron ($\phi=$1). $\dot{E}$ is  normalized to its maximum 
value over the orbital cycle. P=19.3d and e=0.3.
}
\end{SCfigure}

\begin{SCfigure}  
\includegraphics[width=0.65\columnwidth]{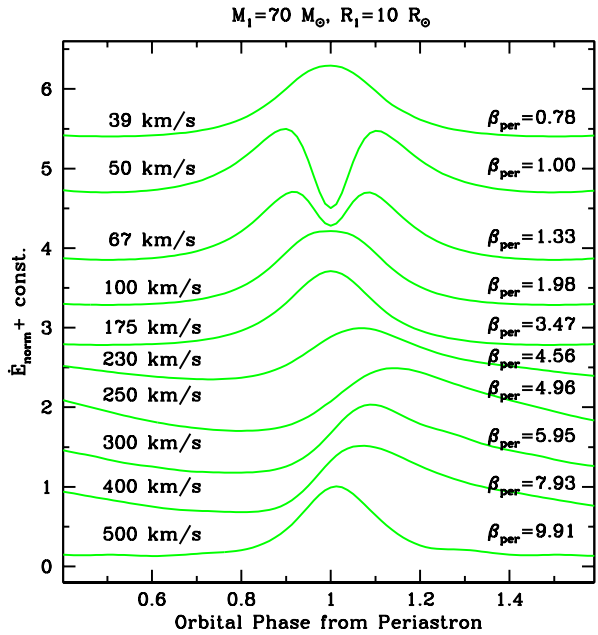}
\caption{$\dot{E}$ plotted as a function of orbital phase for a 70 M$_\odot$, R$_1=$10 R$_\odot$ binary
system with stellar rotation velocities 19--438 km/s.
Maximum $\dot{E}$ values always occur within $\pm$0.1 in phase from periastron.   
}
\end{SCfigure}

\section{Dependence on stellar radius and rotation velocity}

For simplicity, we first consider  a primary star that is  synchronously rotating
at periastron; i.e.,  $\beta_{per}=$1.  Figure 2 illustrates the behavior of energy 
dissipation rates as a function of orbital phase for a 70$+$54 M$_\odot$ binary system
with $P=$19.3 d, $e=$0.3 and values of R$_1=$10--22 R$_\odot$.  For small radii, the
equilibrium tide component dominates. Thus, the trend for increasing  $\dot{E}$ as periastron
is approached reverses just before periastron, and goes to minimum at this phase since 
this is when $\beta_{per}=$1. For stars with larger radii, however, the dynamical tide 
dominates, and $\dot{E}$ presents an oscillatory behavior over the orbital cycle.

Figure 3 illustrates the result of holding constant R$_1=$10R$_\odot$ and varying $v_{rot}$
from sub-synchronous to highly super-synchronous values.  The maximum $\dot{E}$ occurs
close to periastron passage, but not necessarily centered at this orbital phase.   We refer  to the
maximum value attained by $\dot{E}$ over the orbital cycle as $\dot{E}_{Max}$.

Figures 4--6 summarize the behavior of $\dot{E}_{Max}$ as a function of stellar radius and of
rotation velocity for the set of models listed in the Table, and Figure 7 plots the phase
when this maximum is attained.  $\dot{E}_{Max}$ occurs before periastron passage in  
sub-synchronous starsr and after periastron when stars rotate super-synchronously.       


\begin{SCfigure}
\includegraphics[width=0.65\columnwidth]{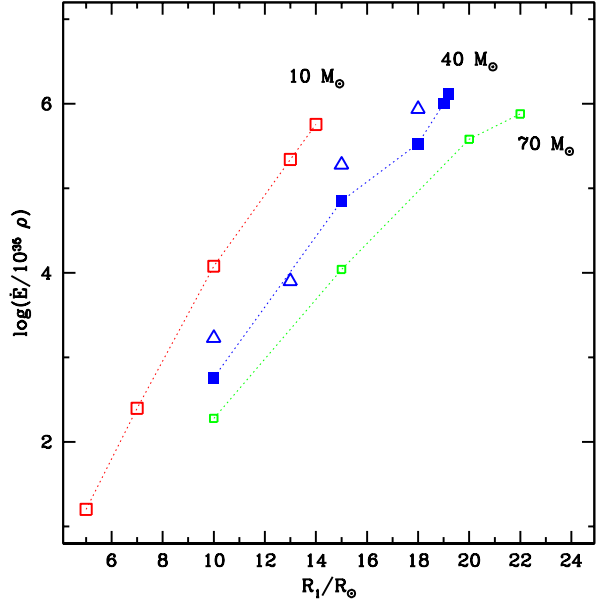}
\caption{$\dot{E}_{Max}$  plotted as a function of 
stellar radius for the $\beta_{per}=$1 calculations listed in the Table.  $\rho$ is
the mean density of the layer where the energy is dissipated.  Squares
correspond to cases with P=19.3, e=0.3 and  triangles to  $P=$100 d, $e=$0.767.  
}
\end{SCfigure}

\begin{SCfigure}
\includegraphics[width=0.65\columnwidth]{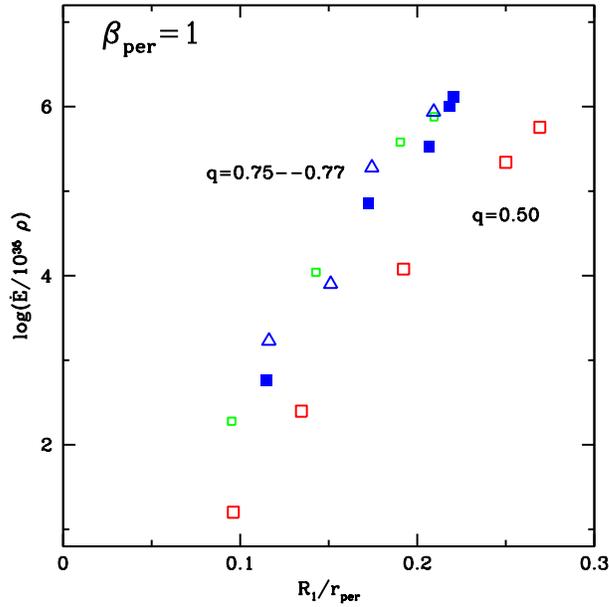}
\caption{Same as previous figure, but here the abscissa is given in
units of stellar radius normalized to the orbital separation at
periastron. The mass-ratio, $q=$M$_2$/M$_1$ is listed.  Colors are as
in Fig. 4.
}
\end{SCfigure}


\begin{SCfigure}  
\includegraphics[width=0.65\columnwidth]{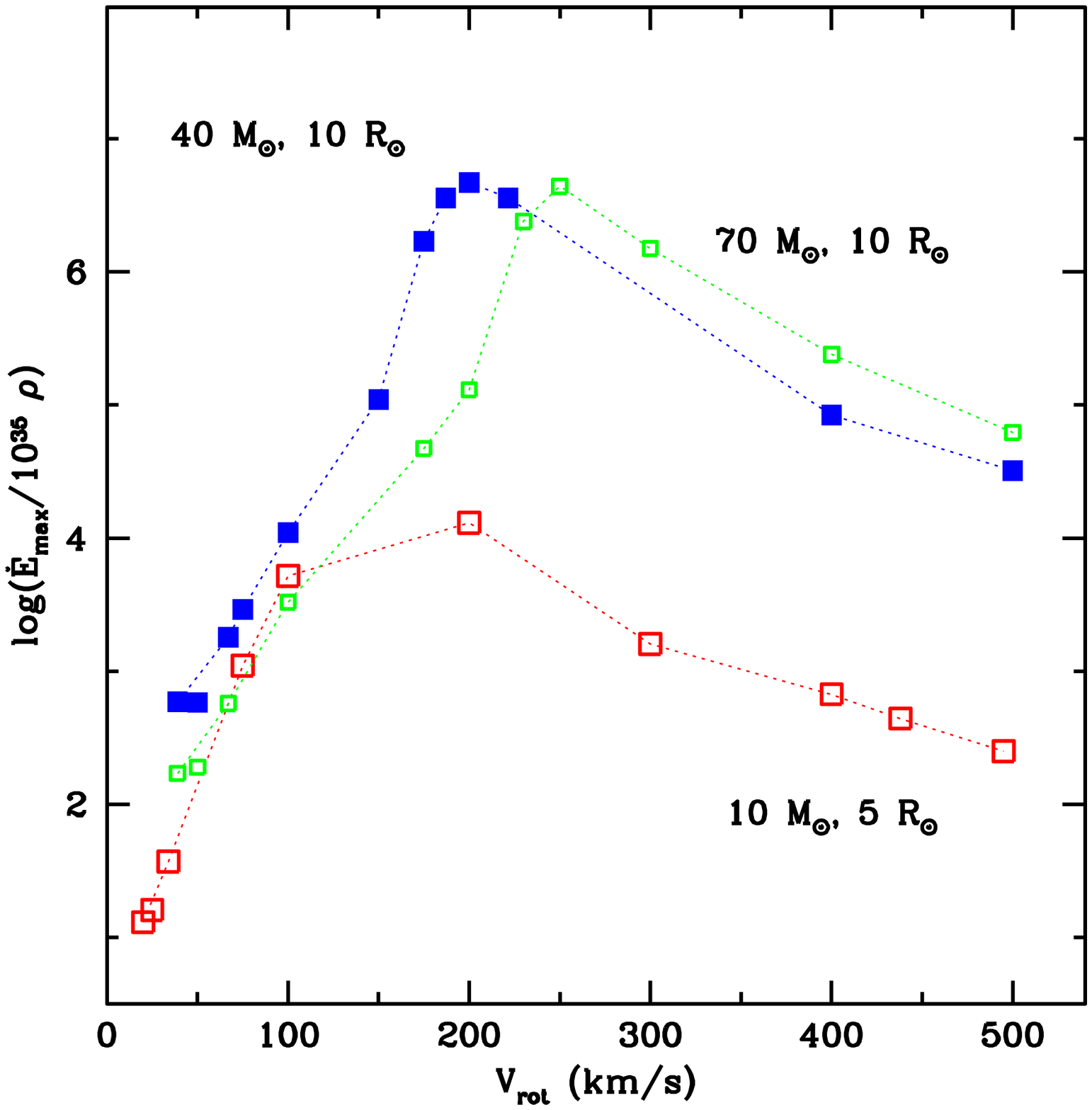}
\caption{$\dot{E}_{Max}$  for  the binary system models
with  P=19.3 d and e=0.3 plotted as a function of equatorial rotation velocities.
Note that ever-increasing rotation velocity does not produce an ever-increasing
$\dot{E}_{Max}$.
}
\end{SCfigure}

\vskip0.3cm
\begin{SCfigure}  
\includegraphics[width=0.65\columnwidth]{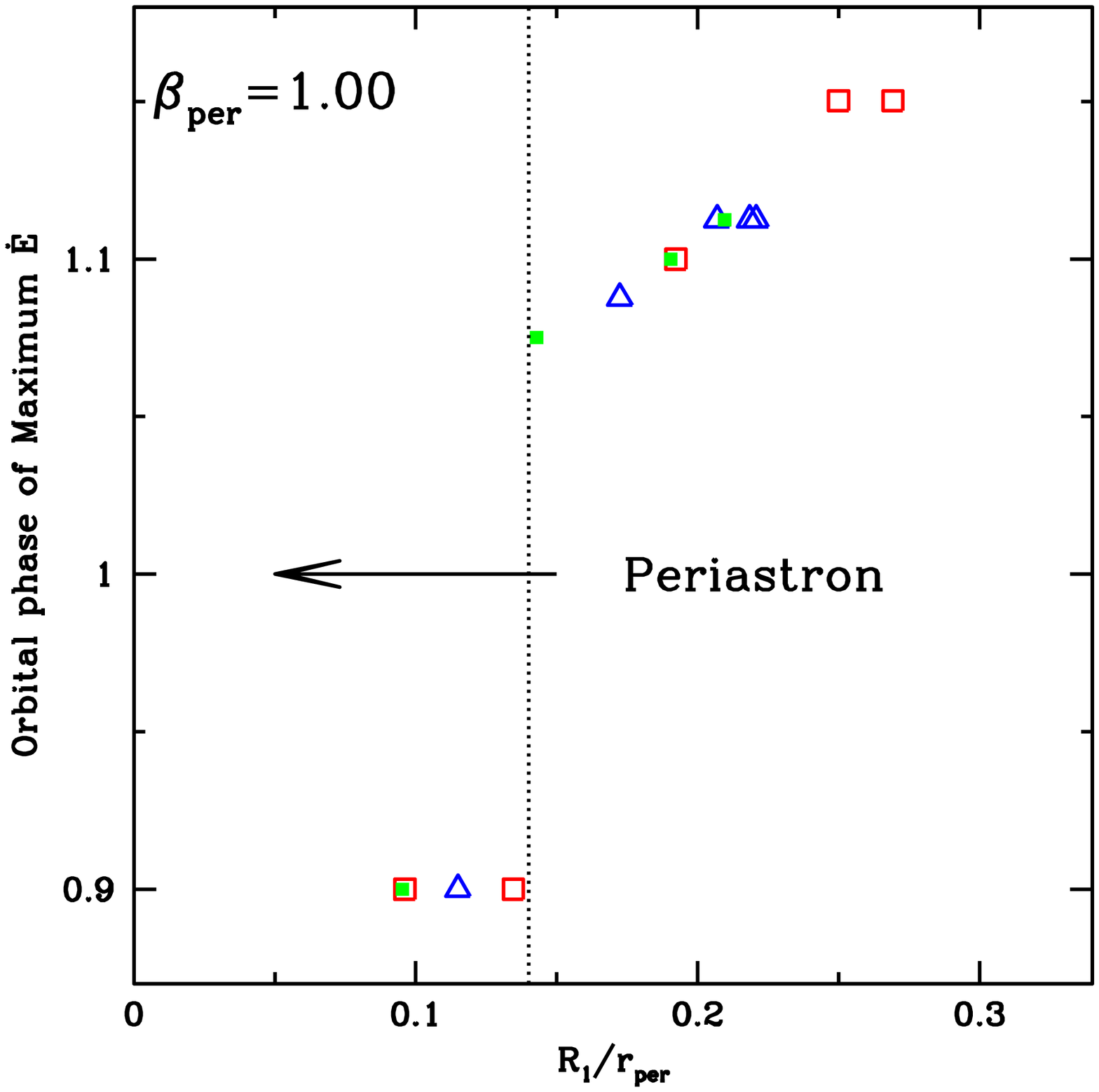}
\caption{Orbital phase when $\dot{E}_{Max}$  occurs for the $\beta_{per}=$1 cases
for M$_1=$ 10 M$_\odot$ (red); 40 M$_\odot$ (blue); 70 M$_\odot$ (green).  
The dotted line shows that for stars with
$R_1$/r$_{per}\leq 0.14$, $\dot{E}_{Max}$ occurs before periastron passage while for stars with larger
relative radii, it occurs after periastron.
}
\end{SCfigure}

\begin{table*}[!h]\centering
\caption{Summary of binary system models}
\begin{tabular}{rrrrrcccc}
\hline
 M$_1$ & M$_2$ & P &  e  &  r$_{per}$ & R$_1$    & V$_{rot}$ & $\beta_0$   & $<\dot{E}_{max}/\rho>$   \\
(M$_\odot$) & (M$_\odot$) & (days)&     & (R$_\odot$)&(R$_\odot$)&(km/s)&           &(10$^{35}$ergs-cm$^3$/s-g)  \\
\hline
    10 &  5      &  19.3 & 0.30 & 52  &  5--14 & 25--71  &  1.000     & 0.16$\times$10$^2$--0.57$\times$10$^6$ \\
    40 & 30      &  19.3 & 0.00 & 74  & 10     & 50      &  1.000     & 0.10$\times$10$^{-2}$                   \\
    40 & 30      &  19.3 & 0.30 & 87  &10--19.2 & 50--97 &  1.000     & 0.58$\times$10$^3$--0.13$\times$10$^7$  \\
    40 & 30      & 100.0 & 0.767& 86  & 10--18 & 59--107 &  1.000     & 0.17$\times$10$^4$--0.87$\times$10$^6$  \\
    40 & 30      & 100.0 & 0.31 &256  & 13     & 13      &  1.000     & 0.10$\times$10$^1$                      \\
    70 & 54      &  19.3 & 0.30 &105  & 10--22 & 50--111 &  1.000     & 0.19$\times$10$^3$--0.76$\times$10$^6$  \\
    10 &  5      &  19.3 & 0.30 & 52  &  5     & 20--495 & 0.79--19.63& 0.13$\times$10$^2$--0.25$\times$10$^3$ \\
    40 & 30      &  19.3 & 0.30 & 87  & 10     & 39--500 & 0.78--2.97 & 0.59$\times$10$^3$--0.47$\times$10$^7$  \\
    40 & 30      & 100.0 & 0.767& 86  & 10     & 39--500 & 0.65--5.08 & 0.28$\times$10$^4$--0.44$\times$10$^7$ \\
   \hline
 \end{tabular}
\end{table*}

\section{Periastron effect in HD 5980 ?}

Figure 8 is a comparison of the recent visual light curve of HD~5980\footnote{Foellmi et al. 2008, RMAA, 44,3}
(kindly provided by P. Massey)
with the $\dot{E}$ values obtained from our model calculation for a 70$+$54 M$_\odot$ binary with 
$\beta_{per}=$1.33 and R$_1=$21~R$_\odot$. The model qualitatively reproduces the behavior after periastron,
suggesting that the tidal flows mechanism may be playing an important role in this system.  Particularly 
interesting also is the  apparent increase in the overall energy flux emitted around periastron ($\phi=$0.15) 
in the wavelength range $\lambda\lambda$1000--10000 \AA.  Illustrated in Figure 9  is the ratio of the 
spectral energy distribution from HST/STIS observations of 1999 and FUSE observations of 2002.  This 
indicates that either there is a significant redistribution in the SED around periastron and/or there is an
added source of energy. Hence, HD 5980 appears to belong in the list of binaries that present a periastron effect.
Furthermore, we speculate that tidal flows may also provide a mechanism to explain the occasional luminous blue variable-like eruptive
behavior.

We suggest the following working scenario: If the erupting star in HD 5980 is in the process of leaving 
the Main Sequence\footnote{as implied by its He/H abundance, Koenigsberger et al. 1998, ApJ, 499, 889.}, its
systematically increasing radius should lead to values of $\dot{E}$ that grow exponentially,
as shown in Figs. 4 and 5. At some point, a critical radius could be reached at which the sum of $\dot{E}$ 
and the intrinsic luminosity of the star becomes super-Eddington, thus producing  an episode of
strongly enhanced mass-loss, which we would observe as an  eruptive event. The frequency of this type of episodes 
might be expected to depend on the expansion rate of the star,
the depth of the layer in which most of the energy dissipation takes place and the amount of mass
that is ejected each time the critical radius is attained.


 \begin{SCfigure}  
\includegraphics[width=0.65\columnwidth]{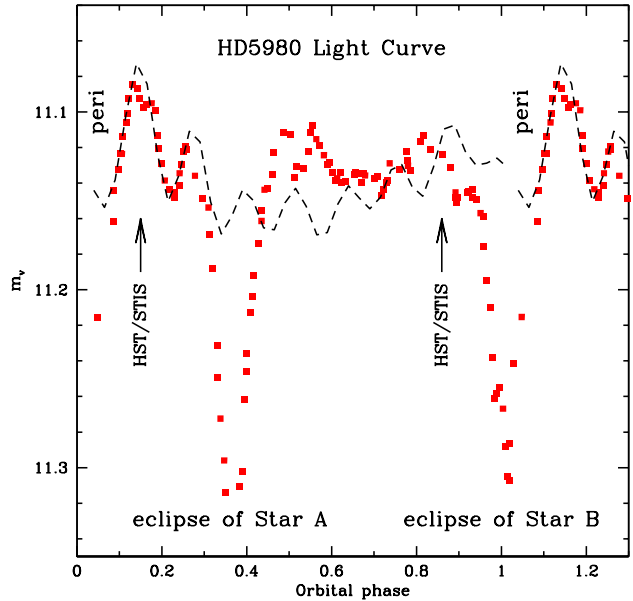}
\caption{V-magnitude light curve of HD 5980 (from
P. Massey) compared  with the tidal energy dissipation curve assuming
R$_A=$21 R$_\odot$ and $\beta_0=$1.33. Times of eclipses and periastron passage
are indicated, as well as phases when the 1999 HST/STIS spectra were obtained.
}
\end{SCfigure}

\begin{SCfigure}  
\includegraphics[width=0.65\columnwidth]{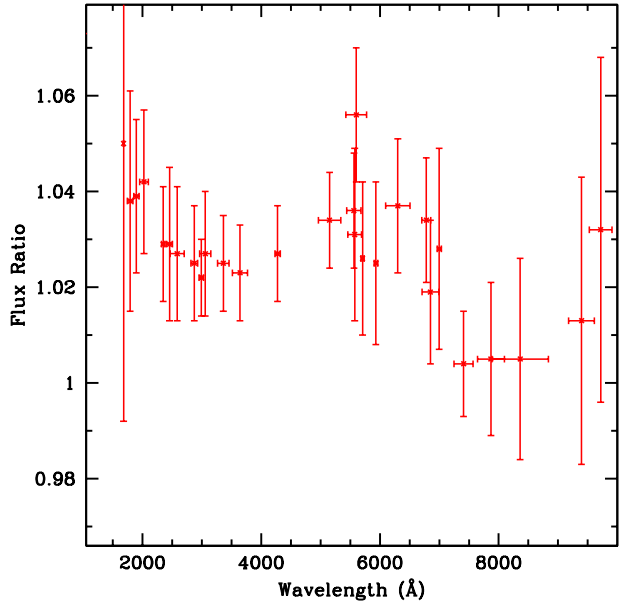}
\caption{Ratio of HST/STIS spectra observed  at orbital phases 0.15 and 0.83 in 1999.  Error bars
are rms of mean for wavelength range indicated by error bars for abscissa.}
\end{SCfigure}

\section{Discussion and future directions}

The  periastron events of $\eta$ Carinae, WR 140 and WR 125 are generally modeled in 
the context of interacting winds theory.\footnote{see Parkin \& Pittard, 2008, MNRAS, 388,104,
and references therein.}  These models have been mostly successful, 
although some discrepancies remain.\footnote{see, for example, Rauw, G., 2008, RMAACS, 33, 59.}   
It is interesting to note that one of the  discrepancies involves the time of maximum X-ray emission, 
which does not always appear at periastron passage as predicted by the wind-wind collision
models.  An explanation for this discrepancy may reside in  that the  mass-loss rate and 
wind velocity structure of the stars does not remain constant over the orbital cycle.  As
shown above, the tidal flows model predicts a significant perturbation of the stellar surface
which, in turn, could significantly affect the stellar wind properties of the stars.  

A further consequence of the tidal flows model  is that the strongest perturbation should occur 
on the larger of the two stars in the binary system.  In the case of the WR+O binary systems, 
the O-star companion generally has the larger stellar radius.  Hence, in a system such as WR 140 
it is conceivable that periastron passage could lead to a significant increase in the  {\em O-star's}
wind momentum.  Assuming this to be the case,  the possibility then exists that the
O-star's wind could dominate the momentum ratio around periastron, thus even inverting the orientation
of the shock cone geometry.  

Finally, it is interesting to note that the action of tidal flows is not distributed symmetrically
over the stellar surface and thus, if the energy dissipated by these motions contributes towards
mass-loss, the additional outflow is expected to depart from  spherical symmetry as well.  Furthermore,
the critical radius could be reached by the star several times during its late stages of evolution,
thereby producing multiple episodes of enhanced mass-loss prior to its demise in the supernova
event.   An interesting problem would be to analyze the morphology of circumstellar structures that may be 
produced from such asymmetric mass-ejections  as well as the manner in which this distribution of 
matter then may affect the expansion of the supernova ejecta.  

In conclusion, the tidal flows  produced in asynchronously rotating  binary systems provide
a mechanism that may explain  a wide variety of observational phenomena ranging from
periodic photospheric line-profile variability to episodic and asymmetric mass-shedding events.
The actual magnitude of the energy dissipation rates depends strongly on the viscosity parameter
that is assumed for  the stellar material and on the depth of the layer where the energy dissipation 
rates are most significant.  A proper assessment of these physical parameters  will help
determine the actual importance of the role being played by   asynchronous rotation and
tidal flows in producing peculiar phenomena observed in binary stars.

\section{Acknowledgements}

This research was supported by UNAM/DGAPA through  PAPIIT grants IN 119205 and IN 106708, and
CONACYT.

}     

\end{document}